\documentstyle[11pt]{article}

\begin{document}

\newcommand{\beq}{\begin{equation}}
\newcommand{\eeq}{\end{equation}}
\newcommand{\bea}{\begin{eqnarray}}
\newcommand{\eea}{\end{eqnarray}}

\newcommand{\chii}{\raise.5ex\hbox{$\chi$}}
\newcommand{\Z}{Z \! \! \! Z}
\newcommand{\R}{I \! \! R}
\newcommand{\N}{I \! \! N}
\newcommand{\C}{I \! \! \! \! C}

\newcommand{\noi}{\noindent}
\newcommand{\vs}{\vspace{5mm}}
\newcommand{\ie}{{${ i.e.\ }$}}
\newcommand{\eg}{{${ e.g.\ }$}}
\newcommand{\ea}{{${ et~al.\ }$}}
\newcommand{\hf}{{\scriptstyle{1 \over 2}}}
\newcommand{\ih}{{\scriptstyle{i \over \hbar}}}
\newcommand{\hi}{{\scriptstyle{ \hbar \over i}}}
\newcommand{\itwoh}{{\scriptstyle{i \over {2\hbar}}}}
\newcommand{\young}{n}
\newcommand{\dbar}[1]{\stackrel{=}{#1}{}}

\newcommand{\deder}[1]{{ 
 {\stackrel{\raise.1ex\hbox{$\leftarrow$}}{\delta^r}   } 
\over {   \delta {#1}}  }}
\newcommand{\dedel}[1]{{ 
 {\stackrel{\lower.3ex \hbox{$\rightarrow$}}{\delta^l}   }
 \over {   \delta {#1}}  }}
\newcommand{\dedetwo}[2]{{    { \delta {#1}} \over {   \delta {#2}}  }}
\newcommand{\dedetre}[3]{{ \left({ \delta {#1}} \over {   \delta {#2}}  
 \right)_{\! \! ({#3})} }}

\newcommand{\papar}[1]{{ 
 {\stackrel{\raise.1ex\hbox{$\leftarrow$}}{\partial^r}   } 
\over {   \partial {#1}}  }}
\newcommand{\papal}[1]{{ 
 {\stackrel{\lower.3ex \hbox{$\rightarrow$}}{\partial^l}   }
 \over {   \partial {#1}}  }}
\newcommand{\papatwo}[2]{{   { \partial {#1}} \over {   \partial {#2}}  }}
\newcommand{\papa}[1]{{  {\partial} \over {\partial {#1}}  }}
\newcommand{\papara}[1]{{ 
 {\stackrel{\raise.1ex\hbox{$\leftarrow$}}{\partial}   } 
\over {   \partial {#1}}  }}

\newcommand{\ddr}[1]{{ 
 {\stackrel{\raise.1ex\hbox{$\leftarrow$}}{\delta^r}   } 
\over {   \delta {#1}}  }}
\newcommand{\ddl}[1]{{ 
 {\stackrel{\lower.3ex \hbox{$\rightarrow$}}{\delta^l}   }
 \over {   \delta {#1}}  }}
\newcommand{\dd}[1]{{  {\delta} \over {\delta {#1}}  }}
\newcommand{\pa}{\partial}
\newcommand{\sokkel}[1]{\!  {\lower 1.5ex \hbox{${\scriptstyle {#1}}$}}}  
\newcommand{\larrow}[1]{\stackrel{\rightarrow}{#1}}
\newcommand{\rarrow}[1]{\stackrel{\leftarrow}{#1}}
\newcommand{\twobyone}[2]{\left(\begin{array}{c}{#1} \cr
                                   {#2} \end{array} \right)}
\newcommand{\twobytwo}[4]{\left[\begin{array}{ccc}{#1}&&{#2} \cr
                                  {#3} && {#4} \end{array} \right]}
\newcommand{\fourbyone}[4]{\left(\begin{array}{c}{#1} \cr{#2} \cr{#3} \cr
                                   {#4} \end{array} \right)}
\newcommand{\sixbyone}[6]{\left(\begin{array}{c}{#1} \cr{#2} \cr{#3} \cr
                                   {#4} \cr{#5} \cr{#6} \end{array} \right)}

\newcommand{\eq}[1]{{(\ref{#1})}}
\newcommand{\Eq}[1]{{eq.~(\ref{#1})}}
\newcommand{\Ref}[1]{{Ref.~\cite{#1}}}
\newcommand{\mb}[1]{{\mbox{${#1}$}}}
\newcommand{\equi}[1]{\stackrel{{#1}}{=}}

\newcommand{\proofbox}{\begin{flushright}
${\,\lower0.9pt\vbox{\hrule \hbox{\vrule
height 0.2 cm \hskip 0.2 cm \vrule height 0.2 cm}\hrule}\,}$
\end{flushright}}


\begin{titlepage}

\title{
\normalsize
\rightline{RU01-01-B}
\rightline{hep-th/0102136}
\vspace{2.5 cm}
\Large\bf Product of Boundary Distributions\\ 
}

\author{{\sc K.~Bering}
\footnote{Email address: {\tt bering@summit.rockefeller.edu, bering@nbi.dk}}
\\
Rockefeller University \\ 
Physics Dept, Box 254 \\ 
1230 York Avenue \\ 
New York, NY, 10021-6399, USA\\
} 

\date{February 2001}

\maketitle
\begin{abstract} 
1) We identify new parameter branches for the ultra-local boundary
Poisson bracket in $d$ spatial dimension with a $(d-1)$-dimensional
spatial boundary. There exist $2^{r(r-1)/2}$ $r$-dimensional 
parameter branches for each $d$-box, $r$-row Young tableau. 
The already known branch (hep-th/9912017) corresponds to a vertical
$1$-column, $d$-box Young tableau. 
2) We consider a local distribution product among the
so-called boundary distributions. The product is required
to respect the associativity and the Leibnitz rule.
We show that the consistency requirements on this product 
correspond to the Jacobi identity conditions for 
the boundary Poisson bracket. In other words, the restrictions 
on forming a boundary Poisson bracket can be related to the more 
fundamental distribution product construction.
3) The definition of the higher functional derivatives is made 
independent of the choice of integral kernel representative for a 
functional.
\end{abstract}

\bigskip

\vspace*{\fill}

\noi
PACS number(s): 02.70.Pt, 11.10.Ef.

\end{titlepage}
\vfill
\newpage

\setcounter{equation}{0}
\section{Introduction}

\noi
It is fair to say that up to now, the Literature 
\cite{soloviev,soloviev2,beringput,solocomp,bering} 
on the boundary Poisson bracket has revolved around the boundary 
Poisson bracket in ``integrated'' form. That is: as a composition 
of two local functionals \mb{F} and \mb{G} into a new local functional
\mb{\{F,G\}}, \ie where all objects are integrated over the full
physical space.

\vs
\noi
We shall in this paper elaborate on a singular, distributional 
formalism for the boundary Poisson bracket. 

\vs
\noi
When addressing this problem, one immediately runs into the well-known 
problem \cite{schwartz1954} of rendering proper sense of 
multiplying very singular distributions together.
Preferably, the distribution product should respect
standard rules of calculus, such as: commutativity, associativity 
and the Leibnitz rule. Perhaps surprisingly, it turns out,
as we shall show in this paper, that a few descriptive rules 
like the above are enough to reproduce, in a one-to-one fashion, 
the family of boundary Poisson bracket solutions, 
of which a $d$-parameter branch was already identified 
in \cite{bering}. From the boundary Poisson bracket point of view,
the restricting principle is the Jacobi identity.

\vs
\noi
Conceptionally, an important question is raised by the 
infinitely many versions of the distribution product that the
above mentioned analysis produces. How do we account for this
infinity? The paradigm will be that a priori they are all equally-good. 
A specific choice relies on external data. That is,
a specific choice of the ``boundary background'',
a specific choice of renormalization scheme for the ``boundary
theory'', etc. We shall simply argue that is no unique, canonical, 
intrisic distribution product. 

\vs
\noi
The main feature of the boundary calculus is the fact that we can 
not freely discard boundary terms, when integrating by part. 
Hence, the adjoint differentiation 
is not just the ordinary differentiation (with a minus in front). 
Curiously, the adjoint differentiation turns out to be 
more fundamental than the differentiation itself.
The logic behind the distributional construction can roughly 
be sketched by the following chain:
\beq
\begin{array}{c}
{\rm Distributions} \cr \Downarrow  \cr
{\rm Adjoint~Differentiation} \cr \Downarrow  \cr 
{\rm Boundary~Distributions} \cr \Downarrow  \cr 
{\rm Product} \cr \Downarrow  \cr 
{\rm Differentiation}~.
\end{array}
\eeq

\vs
\noi
The organization of the paper is as follows: 
We start by reviewing some basic notions in the Schwartzian
\cite{schwartz} formulation of distribution theory, adapted to the
situation with a boundary.  The distribution product itself is 
postponed to Section~\ref{secprod}. We consider the ultra-local case in 
Section~\ref{secul}, where the main results are listed.
Further developments of the general theory are done in Section~\ref{secsupp}
The boundary Poisson bracket is discussed in Section~\ref{secbpb}.
Finally, we consider an alternative basis for the product in 
Section~\ref{secalt}. Conclusions are in Section~\ref{conclus}.
The reader might want to focus on Section~\ref{secprod}-\ref{secul} 
and Section~\ref{secbpb} for a first-time reading.

\setcounter{equation}{0}
\section{Elementary Distribution Theory}

\noi
Let the physical space\footnote{
The reader may substitute the word {\em space} with his favorite term: 
{\em Cauchy surface}, {\em space-time}, {\em world volume}, etc, 
for most of the paper. When addressing the boundary Poisson 
bracket, \mb{\Sigma} will an {\em equal-time-surface}. 
The time variable will be suppressed from the notation.} 
\mb{\Sigma} be a closed subset of \mb{\R^{d}}. Consider the ring 
\mb{C^{\infty}(\R^{d},\C)} of complex-valued functions equipped with
the topology corresponding to the uniform convergence on compact subsets. 
We are not interested in the specific behavior of the functions outside 
the physical space \mb{\Sigma}, so we mode out by the two-sided ideal 
\beq
{\cal I}~\equiv~ \left\{ \left. f \in C^{\infty}(\R^{d},\C) ~\right| 
~ f\left(\Sigma \right)\subseteq \{0\} ~ \right\} 
~\subseteq~C^{\infty}(\R^{d},\C)~.
\eeq
The relevant quotients are two rings and one multiplicative group:
\bea
{\cal E}&\equiv& C^{\infty}\left(\R^{d},\C\right)/{\cal I}~, \cr\cr
{\cal D}&\equiv& C_{c}^{\infty}(\R^{d},\C)/{\cal I} 
~\subseteq ~{\cal E}~,\cr\cr
{\cal E}^{\times}&\equiv& C^{\infty}(\R^{d},\C \backslash \{0\})/{\cal I}
~\subseteq ~{\cal E}~. 
\eea
Elements of the second subquotient ring \mb{{\cal D}} are required to have 
a representative with compact support. We shall loosely refer to this
subring as {\em the test functions with compact support}.
Elements of the multiplicative group 
\mb{C^{\infty}(\R^{d},\C \backslash \{0\})} are required to have 
a non-vanishing representative.

\vs
\noi
The vector space \mb{{\cal D}^{\prime}} of {\em distributions} is here 
defined as the space of functionals 
\mb{u:{\cal E}^{\times} \times {\cal D} \to \C}, which are continuous
and linear wrt.\ the second entry. The second entry carries a test function 
\mb{f \in {\cal D}} while the first entry carries a non-vanishing
volume density \mb{\rho \in {\cal E}^{\times}}.
The support \mb{{\rm supp}(u)} of a distribution \mb{u \in {\cal D}} 
is defined as
\beq
 {\rm supp}(u)~=~\bigcap_{A}\left\{\left. A~{\rm closed}~\subseteq
\R^{d} ~ \right| ~ \forall f \in C^{\infty}\left(\R^{d},\C\right)~:~ 
 f\left(A \right)\subseteq \{0\} ~\Rightarrow~ 
u[{\cal E}^{\times},f]\subseteq\{0\}~ \right\}~. 
\eeq
The ring \mb{{\cal E}} of test functions with not necessarily compact 
support is embedded into the distribution space \mb{{\cal D}^{\prime}} 
via the embedding \mb{i:{\cal E} \to {\cal D}^{\prime}}:
\beq 
i(f)[\rho,g]~\equiv~\int_{\Sigma} \rho~d^{d}x~f~g~,~~~~
\rho \in {\cal E}^{\times}~,~~~~~~~~f \in {\cal E}~,~~~~~~~~g \in {\cal D}~.
\label{iembed}
\eeq
We have that \mb{{\rm supp}(i(f)) \subseteq \Sigma \cap {\rm supp}(f)}. 
Notation: The \mb{i}-map will often not be written explicitly in 
formulas. Hopefully, this will not cause any confusion. 

\vs
\noi
Generally, we have a product 
\mb{{\cal E} \times {\cal D}^{\prime} \to {\cal D}^{\prime}} of a test
function \mb{f} and a distribution \mb{u} via 
\beq
(f \cdot u)[\rho,g]~=~u[\rho,fg]~,~~~~u \in {\cal D}^{\prime}~,~~~~
\rho \in {\cal E}^{\times}~,~~~~f \in {\cal E} ~,~~~~~g \in {\cal D}~.
\label{scalarmult}
\eeq
Note that \mb{{\rm supp}(fu) \subseteq {\rm supp}(f) \cap {\rm supp}(u)}.

\subsection{Adjoint Differentiation}

\noi
Initially, the differential operator 
\mb{\pa_{i}\equiv \papa{x^{i}}:{\cal E} \to {\cal E}}, 
\mb{i=1,\ldots,d}, is only well-defined on smooth functions.
Not every distribution will be differentiable 
when a boundary \mb{\pa\Sigma} is present. Rather, the
distributions are adjointly differentiable.
In detail, the {\em adjoint} differential operator 
\mb{\pa^{\dagger}_{i}:{\cal D}^{\prime} \to {\cal D}^{\prime}} 
acts on a distribution  \mb{u \in {\cal D}^{\prime}} via
\beq
  \pa^{\dagger}_{i}u[\rho,f]~\equiv~u[\rho,\pa_{i}f]~,~~~~
\rho \in {\cal E}^{\times}~,~~~~~~ f \in {\cal D}~.
\eeq
This definition of course mimics
\mb{\int_{\Sigma} \rho~d^{d}x~(\pa^{\dagger}_{i}f)~g
~=~\int_{\Sigma} \rho~d^{d}x~f~\pa_{i}g} for test functions 
\mb{f,g \in {\cal D}}.
We have that 
\mb{{\rm supp}(\pa^{\dagger}_{i}u) \subseteq {\rm supp}(u)}.

\vs
\noi
We immediately get that \mb{[\pa^{\dagger}_{i},\pa^{\dagger}_{j}]u~=~0} 
as a consequence of \mb{[\pa_{i},\pa_{j}]f~=~0}. The commutator 
\beq
  [\pa^{\dagger}_{i},f]~=~-(\pa_{i}f) 
\label{adjderder}
\eeq 
of the adjoint derivative \mb{\pa^{\dagger}_{i}} 
with a test function \mb{f \in {\cal E}} 
is again a test function, as the following short calculation shows:
\bea
[\pa^{\dagger}_{i},f]u[\rho,g]
&=&\pa^{\dagger}_{i} f u[\rho,g]-f \pa^{\dagger}_{i} u[\rho,g]
~=~fu[\rho,\pa_{i}g]-\pa^{\dagger}_{i} u[\rho,fg] \cr \cr
&=&u[\rho,f \pa_{i}g]- u[\rho,\pa_{i}(fg)]
~=~-u[\rho,(\pa_{i}f)g]~=~-(\pa_{i}f)u[\rho,g]~.
\eea
We emphasis that the adjoint derivative 
\mb{\pa^{\dagger}_{i}} does {\em not} satisfy the 
Leibnitz rule, not even among test functions.
Nevertheless, equation \eq{adjderder} can still be viewed as
a Leibnitz-type rule for the adjoint differential operator 
\mb{\pa^{\dagger}_{i}}:
\beq
  \pa^{\dagger}_{i}(fu)~=~-(\pa_{i}f)u+ f\pa^{\dagger}_{i}u~.
\label{adjderder2}
\eeq

\subsection{Boundary Distributions}

\noi
We can now define the main object of this paper:
The vector space \mb{{\cal B}}  of {\em boundary distributions} 
\beq
{\cal B}~\equiv~ \left\{ \left.\sum_{k} f_{k} (\pa^{\dagger})^{k}g_{k}~
\right| {\rm ~All~but~finitely~many~of}~f_{k}, g_{k} \in{\cal E} 
~{\rm are~zero}  \right\} ~\subseteq~ {\cal D}^{\prime}~.
\eeq
We have employed a multi-index notation: For instance, the index 
\beq
k~=~(k_{1}, \ldots, k_{d}) \in \N_{0}^{d} ~,
~~~~~~~\N_{0}~\equiv~\{0,1,2,\ldots\}~,
\label{kayindex}
\eeq
runs over the $d$-dimensional non-negative integers, and
\beq
   (\pa^{\dagger})^{k}~=~(\pa_{1}^{\dagger})^{k_{1}} 
\cdots (\pa_{d}^{\dagger})^{k_{d}}~.
\eeq
By using property \eq{adjderder}, we may write
\beq
  {\cal B}~\subseteq~\bigoplus_{k} {\cal B}_{k}~,~~~~~~~~~~~~ 
{\cal B}_{k}~=~{\cal E} \cdot (\pa^{\dagger})^{k}1~.
\label{boundistdecomp}
\eeq
The adjoint derivatives \mb{\pa^{\dagger}_{i}}, \mb{i=1,\ldots,d}, 
give rise to a $d$-dimensional grading of \mb{{\cal B}}.
The elements \mb{(\pa^{\dagger})^{k}1}, \mb{k \in \N_{0}^{d}} , 
constitute a \mb{{\cal E}}-linear basis for \mb{{\cal B}}, called the 
{\em adjoint basis}. 
Note that the support of the basis elements is the {\em whole} set 
\mb{\Sigma}:
\beq
{\rm supp}\left((\pa^{\dagger})^{k}1\right)~=~\Sigma~.
\eeq 
Here it is crucial that the volume density \mb{\rho} is an
argument of the distributions, so that we can vary \mb{\rho}.
As a consequence, the \mb{{\cal E}}-coefficient function 
inside \mb{{\cal B}_{k}} is unique. (Of course, it is
sufficient to vary \mb{\rho} infinitesimally, for instance
around the constant function \mb{\rho=1}.)

\subsection{Functional and Functional Derivative}
\label{secfunc}

\noi
A {\em functional} \mb{F \in {\cal F}({\cal E}^{\times},\C)} is a map 
\mb{F:{\cal E}^{\times} \to \C} from the group
\mb{ {\cal E}^{\times}} of volume densities to the complex numbers
\mb{\C}, which in general depends on a number of fields 
\mb{\phi^{A}}. We shall not adapt the latter dependence explicitly 
in the notation because it would appear too heavy.
The functional is called {\em differentiable} wrt.\ the fields
\mb{\phi^{A}} if there exist a distributions 
\mb{u_{A}:{\cal E}^{\times} \times {\cal D} \to \C} that satisfies
\beq
 \forall \rho ,  \delta \phi^{B}~:~ 
\sum_{A}u_{A}[\rho,\delta \phi^{A}]~=~\delta F [\rho]
\eeq
for all smooth {\em infinitesimal} variations \mb{\delta \phi^{A}}.
For a differentiable functional \mb{F}, we shall write 
\mb{u_{A}=\delta F/\delta \phi^{A}} and call it the functional 
derivative of \mb{F} wrt.\ \mb{\phi^{A}}.

\vs
\noi
The Lebesque (or the Riemann) integral provides an embedding 
\mb{\int_{\Sigma}: {\cal D} \to {\cal F}({\cal E}^{\times},\C)}, 
of functions \mb{f \in {\cal D}} with compact support
into the functional space 
\mb{{\cal F}({\cal E}^{\times},\C)}:
\beq
\int_{\Sigma} f [\rho]~\equiv~\int_{\Sigma} \rho~d^{d}x~f
~,~~~~\rho \in {\cal E}^{\times}~,~~~~f \in {\cal D}~.
\label{intembed}
\eeq
The integration kernel function \mb{f} is unique -- \ie the 
embedding \mb{\int_{\Sigma}} is injective -- because the volume density 
\mb{\rho} can be varied.
The {\em local} functionals are of this form \mb{F=\int_{\Sigma}f}.
They are differentiable, and their functional derivatives 
are boundary distributions:
\beq
 \dedetwo{F}{\phi^{A}}~=~\sum_{k}(\pa^{\dagger})^{k} P_{A(k)}f 
~\equi{\eq{adjderder}}~
\sum_{k} E_{A(k)}f \cdot (\pa^{\dagger})^{k}1 ~\in~ {\cal B}~.
\label{localdiff}
\eeq
This formula is similar to \cite[Statement 5.2]{soloviev} of Soloviev. Here 
\beq
P_{A(k)}f(x)~\equiv~ \frac{\pa f(x)}{\pa\phi^{A(k)}(x)}
\eeq
and 
\beq
E_{A(k)}f(x)~\equiv~\sum_{m\geq k}\twobyone{m}{k}(-\pa)^{m-k}P_{A(m)}f(x)
\eeq 
denote the higher partial derivatives and the higher Euler-Lagrange 
derivatives, respectively, cf.\ \cite{beringput,olver}.

\subsection{Geometric Consideration}

\noi
One may consider coordinate transformations 
\mb{x^{i}\rightarrow x'^{i'}}. To make this construction geometric, 
one should introduce {\em covariant} derivatives \mb{\pa_{i} \leadsto
D_{i}} by specifying a choice of connection. More generally,
the pertinent multi-index structure becomes non-commutative words 
of a $d$-letter alphabet \cite{beringput}, if the derivatives 
\mb{[D_{i},D_{j}]\neq 0} cease to commute.

\setcounter{equation}{0}
\section{Distribution Product}
\label{secprod}

\subsection{Introduction}

\noi
As is very well known, even without a boundary, \ie \mb{\Sigma=\R^{d}},
there is no truly natural product for two generic distributions 
\mb{u,v \in {\cal D}^{\prime}}. A severe blow against such hopes
is dealt by the Schwartz' Impossibility Result \cite{schwartz1954}: 
It states that one cannot in a satisfactory manner embed the whole 
algebra of {\em continuous} functions \mb{C(\R^{d})} inside a 
differential algebra. In this paper, the term 
{\em differential algebra} denotes an associative and commutative 
algebra equipped with $d$ mutually commutative, linear derivations 
\mb{\pa_{i}}, \mb{i=1,\ldots,d}, that are defined for all algebra 
elements in the algebra and that satisfy the Leibnitz rule.  
In the case with no boundary, Colombeau \cite{colombeau} has 
embedded \mb{{\cal D}^{\prime}} into a differential algebra. As an 
illustrative manifestation of the Schwartz' Impossibility Result at work, 
the product of Columbeau does not coincide with the classical product in 
\mb{C(\R^{d})}. However, it preserves the classical product for smooth
functions \mb{C^{\infty}(\R^{d})}. It would be interesting to know 
if and/or how Columbeau's construction can be adapted to incorporate 
boundaries. 

\vs
\noi
An old idea uses the convolution in the Fourier transformed space 
\cite{hoermander,hoermander2} to define a product. However, we
are concerned, that the important features of the boundary and the notion
of locality would be ``washed away'' by the Fourier transform if one
is not careful.

\subsection{Boundary Poisson Bracket}

\noi
Here we shall take a more modest approach, and only consider a
distribution product within the much smaller class \mb{{\cal B}} of 
boundary distributions. 
Our main motivation and main application 
is to reformulate the boundary Poisson bracket 
\cite{soloviev,beringput,solocomp,bering}
using the boundary distributions. 

\vs
\noi
First step is to make sense of the following product:
\beq
 \{ f, g \} ~=~ 
\dedetwo{\left[\int_{\Sigma}f\right]}{\phi^{A}}
~\omega^{AB}~ \dedetwo{\left[\int_{\Sigma}g\right]}{\phi^{B}}
 ~=~\left(\sum_{k}E_{A(k)}f \cdot (\pa^{\dagger})^{k}1\right) 
\omega^{AB} \left(\sum_{\ell}E_{B(\ell)}g \cdot
 (\pa^{\dagger})^{\ell}1\right)~.
\label{bpb0}
\eeq
In 1993, Soloviev \cite[Rule 5.4]{soloviev} showed that the 
boundary Poisson brackets would have an alternative
description in terms of distribution products. 
He inquired, without pursuing it further, whether a distribution 
product could be given a rigorous meaning.

\subsection{Distribution Product}

\noi
A priori, no canonical choice for such a product is preferred.
A certain choice for the product represents an additional input of
informations into the system. This should be specified in order to 
have a well-posed problem. 
In a fundamental and geometrically formulated theory, like
for instance a complete, non-perturbative formulation of string theory,
the appearance of boundaries themselves, like branes, 
is presumably only a good low energy description. 
By analyzing the product of the boundary distributions, 
we are merely collecting some left-over low energy data of the full  
dynamical boundary theory. As a physicist,
one may view the product as a result of a renormalization without a
regularization. As a mathematician, it is simply going to be a 
definition that depends on external parameters.

\vs
\noi
Let us start by specifying the product on the basis
of \mb{{\cal B}}.  There should exist smooth structure functions
\mb{c^{k,\ell}_{n}(x) \in {\cal E}}, such that
\beq
  (\pa^{\dagger})^{k}1 \cdot
(\pa^{\dagger})^{\ell}1~=~\sum_{n}c^{k,\ell}_{n}~
(\pa^{\dagger})^{n}1~, 
\label{proddef}
\eeq
\ie the algebra \mb{{\cal B}} {\em closes}.
On a technical note, to avoid cluttering up the paper with  
arguments of convergence, we assume that 
\beq
\forall k,\ell~:~  c^{k,\ell}_{n}~=~0
{\rm ~for~all~but~finitely~many} ~n~.
\label{finite}
\eeq 
Next we extend the product to the whole 
\mb{{\cal B}_{k} \times {\cal B}_{\ell}} 
by demanding that the coefficient functions themselves carry no
surprises:
\beq
\left(f_{k} \cdot (\pa^{\dagger})^{k}1\right) 
\cdot \left(g_{\ell} \cdot (\pa^{\dagger})^{\ell}1\right)
 ~=~(f_{k}  g_{\ell}) \cdot 
\left((\pa^{\dagger})^{k}1 \cdot  (\pa^{\dagger})^{\ell}1\right)~.
\eeq
We shall furthermore assume: 
\beq
\begin{array}{rlcccl}
\bullet& {\rm Naturalness}&:& 
u \cdot 1~=~u~=~1 \cdot u~,& {\rm  or}&
c^{k,0}_{n}~=~\delta^{k}_{n}~=~ c^{0,k}_{n}~. \cr\cr
\bullet& {\rm Commutativity}&:& u \cdot v~=~v \cdot u~,&{\rm  or}& 
c^{k,\ell}_{n}~=~ c^{\ell, k}_{n}~.\cr\cr
\bullet& {\rm Associativity}&:& 
(u \cdot v) \cdot w~=~u \cdot (v\cdot w)~,&{\rm  or}&
\sum_{n} c^{k,\ell}_{n}~ c^{n, m}_{p}
~=~\sum_{n}c^{k, n}_{p}~ c^{\ell, m}_{n}~.
\end{array}
\eeq
To ensure that the boundary Poisson bracket (See Section~\ref{secbpb}) 
does not change the bulk properties, one can impose the decoupling 
requirement:
\beq
\bullet~ {\rm Decoupling}~:~~~~~ 
u \perp 1 ~~\vee~~ v \perp 1 ~~~\Rightarrow~~~ u\cdot v \perp 1 ~,~~~~~~~ ~
{\rm  or}~~~~~~~~~~c^{k,\ell}_{0}~=~\delta^{k}_{0}~\delta^{\ell}_{0}~.
\eeq

\vs
\noi
Strictly speaking, we cannot exclude the possibility that going
to a bigger algebra than the boundary distribution algebra \mb{{\cal
B}} might kill some of the solutions by additional requirements. 
However, this does not jeopardize the \mb{{\cal B}} product
construction as an isolated construction. 
It rests in itself. In fact, rather we predict that one can construct
deformed versions of the Colombeau algebra.

\subsection{Differentiation}

\noi
Given the multiplication in \mb{{\cal B}}, there now is a canonical 
choice for a differential structure \mb{\pa_{i}} in \mb{{\cal B}}. 
This consists in requiring that the Leibnitz-type relation \eq{adjderder} 
which previously only applied to smooth functions should continue 
to hold for all boundary distributions \mb{u \in {\cal B}}:
\beq
\pa_{i} u~\equiv~u \cdot \pa^{\dagger}_{i}1 - \pa^{\dagger}_{i}u~.
\label{defder}
\eeq
Note how the existence of the product is vitale for this definition.
One may easily check that \mb{\pa_{j}i(f)=i(\pa_{j}f)} for smooth 
functions \mb{f\in {\cal E}}, cf.\ \eq{iembed}, and that
this definition \eq{defder} respects the Leibnitz rule wrt.\ 
the standard product \mb{\cdot:{\cal E} \times {\cal B} \to {\cal B}},
cf.\ \eq{scalarmult}.
We shall also impose the Leibnitz rule for the full product 
\mb{\cdot:{\cal B} \times {\cal B} \to {\cal B}}:
\beq
   \pa_{i}(u \cdot v)~=~ (\pa_{i} u) \cdot v~+~u \cdot \pa_{i} v~.
\label{sleibner}
\eeq
The Leibnitz rule \eq{sleibner} leads to a tower of quadratic consistency 
relations among the structure functions:
\beq
  c^{k+e_{i},\ell}_{n}+ c^{k,\ell+e_{i}}_{n}
 -\sum_{p}c^{k,\ell}_{p}~c^{p,e_{i}}_{n}
~=~ c^{k,\ell}_{n-e_{i}}-\pa_{i} c^{k,\ell}_{n}~.
\label{qualeibcond}
\eeq
Here \mb{e_{i}\equiv(0,\ldots,0,1,0,\ldots,0)} is the $i$'th unit vector
of the index lattice. We shall extensively investigate this quadratic 
Leibnitz condition in the following two Sections~\ref{secul}-\ref{secsupp}.
With the implimentation of the Leibnitz rule, we may refer to 
\mb{\pa_{i}} as a (first order) linear derivation, while
\mb{\pa^{\dagger}_{i}} is a (first order) affine differential operator.
Hence their commutator becomes a (zeroth order) left multiplication 
operator: 
\beq \left\{
\begin{array}{rcl}
[\pa^{\dagger}_{i},\pa^{\dagger}_{j}]&=&0~, \cr\cr
[\pa_{i},\pa_{j}]&\equi{\eq{sleibner}}&0~, \cr\cr
[\pa_{i},\pa^{\dagger}_{j}]&\equi{\eq{sleibner}}&
\pa_{i}\pa^{\dagger}_{j}1
~\equiv~\pa^{\dagger}_{i}1\cdot\pa^{\dagger}_{j}1
-\pa^{\dagger}_{i}\pa^{\dagger}_{j}1 ~.
\end{array} \right.
\label{derivalg}
\eeq

\setcounter{equation}{0}
\section{Ultra-Local Product}
\label{secul}

\subsection{Definition}
\label{secultraadj}

\noi
The product is called {\em ultra-local}, if the \mb{c^{k,\ell}_{n}} 
coefficient functions are of the form 
\beq
  c^{k,\ell}_{n}~=~ c^{k,\ell}~\delta^{k+\ell}_{n}~.
\eeq  
In this distinguished case the structure constants \mb{c^{k,\ell}_{n}} 
are dimensionless, and the finiteness assumption \eq{finite} is
automatically satisfied. The decoupling requirement follows from 
the naturalness assumption. Moreover, the other assumptions reduces to:
\beq
\begin{array}{rlcl}
\bullet& {\rm Naturalness}&:&
c^{k,0}~=~1~=~ c^{0,k}~. \cr\cr
\bullet& {\rm Commutativity}&:& 
c^{k,\ell}~=~ c^{\ell, k}~.\cr\cr
\bullet& {\rm Associativity}&:& 
 c^{k,\ell}~ c^{k+\ell, m}
~=~c^{k,\ell+ m}~ c^{\ell, m}~.\cr\cr
\bullet& {\rm Leibnitz~rule}&:&
 c^{k+e_{i},\ell}+ c^{k,\ell+e_{i}}
 -c^{k,\ell}~c^{k+\ell,e_{i}}~=~ c^{k,\ell}~~~~
\wedge ~~~~\pa_{i} c^{k,\ell}_{n}~=~0~.
\end{array}
\eeq
Note that the Leibnitz rule has split into two conditions. The last 
condition says that there will only be locally constant solutions, 
\ie they are constant on each connected component of \mb{\Sigma}.
Let us restrict ourselves to one of the connected components, so we
can speak of a constant solution. Then the commutativity and the 
associativity conditions define a \mb{2}-cocycle in the Abelian 
multiplicative monoid \mb{(\C,\cdot)} of complex numbers. 
(A {\em monoid} is a semi-group with a neutral element. 
Contrary to the notion of a group, elements are not assumed 
to have an inverse element.)

\subsection{Ultra-Local Analysis}
\label{secultraana}

\noi
The generic solutions are \mb{2}-coboundaries:
\beq 
\exists \{b^{k} \}_{k} ~\subseteq~ \C \cup \{\infty\}~:~
c^{k,\ell}~=~\frac{b^{k}~b^{\ell}}{b^{k+\ell}}~,~~~~~~~~b^{k=0}~=~1~,
\eeq
that satisfies the first of the Leibnitz conditions. By use of the 
exactness, it can be cast into:
\beq
\frac{1}{c^{k,e_{i}}}+\frac{1}{c^{\ell,e_{i}}}
~=~\frac{1}{c^{k+\ell,e_{i}}}+1~.
\eeq
It follows that there exists constants 
\mb{s_{ij} ~\in~ \C \cup \{\infty\}}, \mb{i,j=1,\ldots,d}, such that 
\beq
 \frac{b^{k+e_{i}}}{b^{k}~b^{e_{i}}} ~=~ 
\frac{1}{c^{k,e_{i}}}~=~
1+\sum_{j=1}^{d}\frac{k_{j}}{s_{ij}}~.
\eeq
Let us calculate the first few lattice vectors in terms of the 
basis elements:
\bea 
b^{e_{i}+e_{j}}&=&b^{e_{i}}~b^{e_{j}}~(1+\frac{1}{s_{ij}})~, \cr\cr
b^{e_{i}+e_{j}+e_{k}}&=&b^{e_{i}}~b^{e_{j}}~b^{e_{k}}~
(1+\frac{1}{s_{ij}})(1+\frac{1}{s_{ik}}+\frac{1}{s_{jk}})~. 
\eea
Enforcing the symmetry in the first formula and the associativity 
in the last formula we are lead to consider the following algebraic 
variety in \mb{\left( \C \cup \{\infty\} \right)^{d^2}}:
\beq
 s_{ij}~=~s_{ji}  ~~~\wedge~~~s_{ij}~( s_{ik}- s_{jk})~=~0~.
\label{bass}
\eeq
(no sums.) It turns out that there are no further requirements and that we
can associate a solution to each parameter value \mb{\{s_{ij}\}_{ij}}
on this algebraic variety. The latter requirement in \eq{bass} has the
solutions
\beq
\begin{array}{rcccccl}
&s_{ij}~=~0 &\vee&s_{ij}~=~\infty &\vee& 
\forall k=1,\ldots, d~:~ s_{ik}~=~s_{jk}~.& \cr
\end{array}
\eeq
In words, if \mb{s_{ij}\in  \C \backslash \{0\}}, then 
the \mb{i}'th and the \mb{j}'th row (and column) are pairwise equal.
This inspires an equivalence relation ``\mb{\sim}'' in \mb{\{1,\ldots,d\}}:
\beq
 i \sim j ~~~\stackrel{{\rm def.}}{\Leftrightarrow}
~~~i=j~~\vee~~s_{ij}\in  \C \backslash \{0\}~.
\eeq
(It is reflexive by construction. The symmetry and the transitivity 
follow from \eq{bass}.) Let us call the number of equivalence 
classes the {\em rank}, and denote it \mb{r}. The cardinality 
\mb{\young_{i}} of the equivalence classes has the sum 
\mb{\sum_{i=1}^{r}\young_{i}=d}. Clearly, the equivalence classes can be 
mapped into a $d$-box, $r$-row Young tableau of row lengths \mb{(\young_{1},
\ldots, \young_{r})}, where 
\mb{\young_{1} \geq \ldots \geq \young_{r} \geq 1}. 
(We remind the reader that a {\em Young tableau} is a
decorated {\em Young diagram}, where the decoration consists in a
numbering \mb{1,\ldots,d} of the boxes. The boxes within the same row
and rows of equal lengths can still freely be permuted.) 
We note that after the row (and column) reduction ``\mb{\sim}'', 
the \mb{s_{ij}} matrix is an
\mb{r \times r} symmetric matrix with \mb{\frac{r(r-1)}{2}} free
off-diagonal elements \mb{s_{ij}=0} or \mb{s_{ij}=\infty}, that can take
\mb{2^{\frac{r(r-1)}{2}}} different combinations. 

\vs
\noi
When all diagonal elements 
\mb{s_{i}\equiv s_{ii} \in \C \backslash \{0\}}, we are on a
$r$-dimensional branch of the algebraic variety. Branch points have 
more than one of the diagonal elements \mb{s_{i}} equal to 
\mb{s_{i}=0} or \mb{s_{i}=\infty}.
Branch points only exist in higher dimensions \mb{d>1}.

\subsection{Ultra-Local Results}

\noi
Our main result is that we can write the generic solution as
a product of two factors \mb{(I)} and \mb{(II)}:
\bea
  c^{k,\ell}&=&c_{(I)}^{k,\ell}~c_{(II)}^{k,\ell}~, \cr\cr
 c_{(I)}^{k,\ell}&=&  \frac{(s)_{i(k)}(s)_{i(\ell)}}{(s)_{i(k+\ell)}}~=~
 \frac{\Gamma(i(k)\!+\!s)~\Gamma(i(\ell)\!+\!s)}
{\Gamma(i(k\!+\!\ell)\!+\!s)~\Gamma(s)} \cr\cr 
&=&\frac{B(i(k)\!+\!s,i(\ell)\!+\!s)}{B(i(k\!+\!\ell)\!+\!s,s)}
~=~\prod_{i=0}^{i(k)-1}\prod_{j=0}^{i(\ell)-1}\frac{i+j+s}{i+j+1+s}~,\cr\cr
c_{(II)}^{k,\ell}&=&\prod_{\begin{array}{c}
 i,j \cr i \neq j \cr s_{ij}=0\end{array}} 
\delta_{0}^{{\rm min}(k_{i}+\ell_{i},k_{j}+\ell_{j})}~,
\label{mainresult}
\eea
for arbitrary diagonal parameters 
\mb{s\in \left((\C \cup \{\infty\}) \backslash (-\N)\right)^{r}} 
on $r$ copies of the Riemann sphere, \mb{r=1,\ldots,d}.  
Here \mb{(s)_{n}=\Gamma(s+n)/\Gamma(s)} 
is the Pochhammer symbol in $r$ dimensions and 
\mb{\Gamma(s)\equiv\prod_{i=1}^{r}\Gamma(s_{i})}. 
The linear map \mb{i:\C^{d}\to\C^{r}} maps the basis vectors \mb{e_{i}} to
 the basis vectors \mb{i(e_{j})~=~e_{j/\sim}}. That is:
\beq
i(e_{j})~=~i(e_{k}) ~~\Leftrightarrow~~ j \sim k~.
\eeq

\vs
\noi
The solutions \eq{mainresult} are precisely the allowed coefficients 
for the ultra-local boundary Poisson bracket. As we shall see, this 
is by no means a coincidence. It merely reflects a deep correspondance
of notions and properties between the two kinds of boundary calculi 
that will be further illuminated in Section~\ref{secbpb}.
In \cite{bering} we found one $d$-dimensional branch.

\subsection{Examples}

\noi
As an example, consider the vertical, full-rank Young diagram 
\mb{(\young_{1}\!=\!1,\ldots,\young_{d}\!=\!1)}. 
If all the off-diagonal parameters \mb{s_{ij}=\infty} blow up, 
the solution reads:
\beq
 c^{k,\ell}~=~\frac{(s)_{k}(s)_{\ell}}{(s)_{k+\ell}}~=~
\frac{\Gamma(k\!+\!s)\Gamma(\ell\!+\!s)}
{\Gamma(k\!+\!\ell\!+\!s)\Gamma(s)}~.
\eeq
This is the $d$-dimensional branch found in \cite{bering}.

\vs
\noi
If all the off-diagonal parameters \mb{s_{ij}=0} vanish instead, 
the solution reads:
\beq
 c^{k,\ell}~=~ \left\{ 
\begin{array}{cccl}
\frac{(s_{i})_{k_{i}}(s_{i})_{\ell_{i}}}
{(s_{i})_{k_{i}+\ell_{i}}}&=&
\frac{\Gamma(k_{i}\!+\!s_{i})\Gamma(\ell_{i}\!+\!s_{i})}
{\Gamma(k_{i}\!+\!\ell_{i}\!+\!s_{i})\Gamma(s_{i})}
& {\rm if}~\exists i=1,\ldots,d~:~
 k \parallel e_{i} \parallel \ell ~,\cr\cr
0&&& {\rm otherwise}~. 
\end{array} \right.
\eeq

\vs
\noi
As a third example, consider the horizontal, maximally symmetric, 
rank-$1$ Young diagram \mb{(\young_{1}\!=\!d)}, 
where all the parameters \mb{s_{ij}=s} are equal. 
In this case, the solution is simply:
\beq
 c^{k,\ell}~=~\frac{(s)_{|k|}(s)_{|\ell|}}{(s)_{|k|+|\ell|}}~=~
\frac{\Gamma(|k|\!+\!s)\Gamma(|\ell|\!+\!s)}
{\Gamma(|k|\!+\!|\ell|\!+\!s)\Gamma(s)}~,
\eeq
where \mb{|k|~\equiv~\sum_{i=1}^{d}k_{i}}.
This distinguished $1$-dim branch has contrary to all other branches 
an alternative \mb{\Delta^{(n)}} basis, as we shall see in 
Section~\ref{secdeltaopinv}.

\setcounter{equation}{0}
\section{Supplementary Formalism}
\label{secsupp}

\subsection{Integration of Distributions}
\label{secintdist}

\noi
Note that when a distribution has compact support, its test functions do 
not also have to have compact support, understood in the appropriate sense.
Hence, we may extend the definition of integration 
\mb{\int_{\Sigma}:{\cal D} \to {\cal F}({\cal E}^{\times},\C)}, 
cf.\ \eq{intembed}, 
to an arbitrary boundary distribution 
\mb{u \in {\cal B}_{c}} with compact support, simply by:
\beq
 \int_{\Sigma} \rho~d^{d}x ~u~ \equiv~ 
\int_{\Sigma}u[\rho]  ~=~u[\rho,1]~.
\label{intembed2}
\eeq
The integral \eq{intembed2} satisfies
\beq
\int_{\Sigma} \rho~d^{d}x 
\left( u \cdot \pa^{\dagger}_{i}1 -\pa_{i} u \right)
~=~\int_{\Sigma}\rho~d^{d}x ~\pa^{\dagger}_{i}u~=~ 
\pa^{\dagger}_{i}u[\rho,1]~=~u[\rho,\pa_{i}1]~=~u[\rho,0]~=~0~.
\label{leiblemma}
\eeq
This together with the Leibnitz rule yields
\mb{\int_{\Sigma} \rho~d^{d}x~(\pa^{\dagger}_{i}u)\cdot v
~=~\int_{\Sigma} \rho~d^{d}x~u\cdot\pa_{i}v} for boundary distributions 
\mb{u,v \in {\cal B}_{c}}.
We see from the second expression in \eq{leiblemma}
that even with a boundary \mb{\pa \Sigma} present, the integral of a
total {\em adjoint} derivative vanishes. The opposite is also true,
as will become clear below:
\beq
\int_{\Sigma}u~=~0~~~~~~ \Leftrightarrow ~~~~~~
\exists~ u^{1},\ldots, u^{d}~\in~ {\cal B}_{c}~:~
u~=~\sum_{i=1}^{d}\pa^{\dagger}_{i}u^{i}~.
\label{totaladj}
\eeq
(Here it is important that \mb{\rho} is free.)

\vs
\noi
More explicitly, given an arbitrary boundary distribution
\mb{u~=~\sum_{k}f_{k} \cdot (\pa^{\dagger})^{k}1 ~\in~ {\cal B}_{c}},
with finitely many smooth coefficient functions \mb{f_{k}\in {\cal D}},
there exists a {\em unique} smooth function 
\beq
f~=~\sum_{k}\pa^{k}f_{k}~\in~{\cal D}~, 
\eeq
which has the same integral: 
\bea
\int_{\Sigma}u&=& \int_{\Sigma}\sum_{k}f_{k} \cdot
(\pa^{\dagger})^{k}1~=~\sum_{k}(\pa^{\dagger})^{k}1[\cdot~,f_{k}] 
~=~\int_{\Sigma}\sum_{k}\pa^{k}f_{k}\cr\cr
&=&\int_{\Sigma}f~
\equi{(\ref{iembed})}~i(f)[\cdot~,1]~=~\int_{\Sigma}i(f) ~,
~~~~~~~~~~~~~~~~~f_{k} \in {\cal D}~.
\label{disttof}
\eea
See Soloviev \cite[p.2]{soloviev2} for a related formula. 
The non-injective linear map \mb{u \mapsto \int_{\Sigma}u} 
has the following kernel:
\beq
  \int_{\Sigma}u~=~0~~~~~~\Leftrightarrow~~~~~~f~=~0.
\eeq
We can see this in a more elaborated manner by rewriting \mb{u} as
\bea
 u&=&\sum_{k}f_{k} \cdot (\pa^{\dagger})^{k}1~\equi{\eq{adjderder}}~
\sum_{k}(\pa^{\dagger})^{k} \sum_{m\geq k} \twobyone{m}{k}\pa^{m-k}f_{m}
\cr\cr &=& f + ~({\rm total~ adjoint~ derivative~terms})~.
\eea
This also establishes a proof of \eq{totaladj}.
We shall sometimes refer to the unique smooth function \mb{f} as 
\mb{f\equiv C^{\infty}(u)}.

\vs
\noi 
The functional \mb{\int_{\Sigma}u} is differentiable as a consequence 
of \eq{disttof} and our previous discussion in Section~\ref{secfunc}
of the \mb{C^{\infty}} case.
It follows from straightforward manipulations,
that \eq{localdiff} is valid for distributions as well:
\bea
  \dedetwo{\left[\int_{\Sigma}u \right]}{\phi^{A}}
&=& \dedetwo{\left[\int_{\Sigma}f \right]}{\phi^{A}}
~=~\sum_{k} E_{A(k)}f \cdot (\pa^{\dagger})^{k}1
~=~ \sum_{k}(\pa^{\dagger})^{k} P_{A(k)}f \cr\cr
&=&\sum_{k}(\pa^{\dagger})^{k} P_{A(k)}u
~=~\sum_{k} E_{A(k)}u \cdot (\pa^{\dagger})^{k}1 ~,
\label{localdiff2}
\eea  
where we have independently extended the definition \mb{P_{A(k)}u\equiv 
\sum_{n}P_{A(k)}f_{n}\cdot (\pa^{\dagger})^{n}1}
and 
\beq
E_{A(k)}u~\equiv~\sum_{m\geq k}
\twobyone{m}{k}(-\pa)^{m-k}P_{A(m)}u
\eeq
to an arbitrary boundary distribution \mb{u \in {\cal B}}.

\vs
\noi
The functionals can in principle also depend on the mixed derivatives 
\mb{P(\pa,\pa^{\dagger})\phi^{A}} of the field variables
\mb{\phi^{A}}, where \mb{P(\pa,\pa^{\dagger})} is a polynomial in the
non-commutative \mb{\pa_{i}} and \mb{\pa^{\dagger}_{i}}.
However, we shall for simplicity assume thoughout this paper that
adjoint derivatives \mb{\pa^{\dagger}_{i}} acting on the fundamental field
variables \mb{\phi^{A}} have been reexpressed in terms of the ordinary 
derivatives and multiplication operators by repeated use of \eq{defder}.
Let us also note for later that
\beq
 [\pa^{\dagger}_{i}, P_{A(k)}]
~=~[ P_{A(k)},\pa_{i}]~=~P_{A(k-e_{i})}~.
\label{heisalg}
\eeq

\subsection{Generating Functions and Fourier Transform}
\label{secgen}

\noi
In order to carry out the analysis most efficiently, it
is convenient to resum the distributions in a generating series
\bea
 e^{\pa^{\dagger}y}u&\equiv&\sum_{k} \frac{y^{k}}{k!} 
~(\pa^{\dagger})^{k}u~, \cr\cr 
 e^{\pa y}u&\equiv&\sum_{k} \frac{y^{k}}{k!} ~\pa^{k} u 
~=~ \frac{e^{-\pa^{\dagger}y}u}{e^{-\pa^{\dagger}y}1}~.
\label{genderdef}
\eea
The last expression should be understood by a formal Taylor
expansion around \mb{y=0}. Then it merely reproduces the previous 
definition \eq{defder} of differentiation \mb{\pa_{i}} and its
iterated consequences (assuming the Leibnitz rule). The 
Leibnitz-type rule \eq{adjderder2} for the adjoint differential operator 
\mb{\pa^{\dagger}_{i}} is
\beq
e^{\pa^{\dagger}y}(f \cdot u)~=~ 
e^{-\pa y}f \cdot e^{\pa^{\dagger}y}u~,~~~~~~~~~~~~~~
f \cdot e^{\pa^{\dagger}y}u
~=~ e^{\pa^{\dagger}y}(e^{\pa y}f \cdot u)~,
\eeq
while the Leibnitz rule \eq{sleibner} for the differential operator 
\mb{\pa_{i}} becomes
\beq
e^{\pa y}(u \cdot v)~=~ 
e^{\pa y}u \cdot e^{\pa y}v~,
\eeq 
or equivalently,
\beq
 e^{\pa^{\dagger}y}(u \cdot v)~=~ 
\frac{e^{\pa^{\dagger}y}u \cdot 
e^{\pa^{\dagger}y}v }{e^{\pa^{\dagger}y}1}~.
\eeq
We have the following ``affine Leibnitz rule'':
\beq
e^{u_{A}+u_{B}+\pa y}(u\cdot v)
~=~e^{u_{A}+\pa y}u\cdot e^{u_{B}+\pa y}v~,~~~~~~~~~~
u_{A},u_{B},u,v~\in~ {\cal B}~.
\eeq
The interplay \eq{derivalg} between  \mb{\pa^{\dagger}_{i}} and 
\mb{\pa_{i}} reads:
\beq
e^{-\pa y}e^{\pa^{\dagger}\bar{y}}u
~=~\frac{e^{-\pa y}e^{\pa^{\dagger}\bar{y}}1}
{e^{\pa^{\dagger} \bar{y}}1}
e^{\pa^{\dagger}\bar{y}}e^{-\pa y}u
~=~\frac{e^{\pa^{\dagger}(y+\bar{y})}1}
{e^{\pa^{\dagger}y}1\cdot e^{\pa^{\dagger} \bar{y}}1}
e^{\pa^{\dagger}\bar{y}}e^{-\pa y}u~.
\eeq

\vs
\noi
The product \eq{proddef} is summed up as
\beq
T(y_{A},y_{B},q)~\equiv~\sum_{k,\ell,n} 
\frac{y_{A}^{k}}{k!}~\frac{y_{B}^{\ell}}{\ell!}~
c^{k,\ell}_{n} ~q^{n}~.
\eeq
It has the Fourier transform\footnote{
Notation: A $q$-integration 
\mb{\int dq ~\leadsto~ \int_{-\infty}^{+\infty} dq} is always taken to
be along the real axis \mb{\R}, while a $y$-integration \mb{\int dy
~\leadsto~ \int_{-i \infty}^{+i \infty} \frac{dy}{2\pi i}}, is along
the imaginary axis \mb{i\R}, and normalized with a factor \mb{2\pi
i}. }\footnote{We have shuffled the order of the
arguments \mb{T(y,y_{A},y_{B}) \leadsto T(y_{A},y_{B},y) } 
compared with a similar notation used in \cite{bering}.}
\beq
T(y_{A},y_{B},y)~\equiv~
\int \! d^{d}q~e^{-q y}~T(y_{A},y_{B},q)~.
\eeq
The boundary algebra then reads
\beq
\begin{array}{rlcl}
\bullet& {\rm Closure}&:& 
 e^{\pa^{\dagger}y_{A}}1 \cdot  e^{\pa^{\dagger}y_{B}}1~=~
\int \! d^{d}y~T(y_{A},y_{B},y)~ e^{\pa^{\dagger}y}1~.  \cr\cr
\bullet& {\rm Naturalness}&:& 
 T(y_{A},0,y)~=~ \delta^{d}(y_{A}\!-\!y)~=~ T(0,y_{A},y)~. \cr\cr
\bullet& {\rm Commutativity}&:& T(y_{A},y_{B},y)~=~T(y_{B},y_{A},y)~.\cr\cr
\bullet& {\rm Associativity}&:&
\int \! d^{d} \tilde{y}~T(y_{A},y_{B},\tilde{y})~T(\tilde{y},y_{C},y)
~=~\int \! d^{d} \tilde{y}~T(y_{A},\tilde{y},y)~T(y_{B},y_{C},\tilde{y})~.
\cr\cr
\bullet& {\rm Leibnitz~ Rule}&:&
 \int \! d^{d} \tilde{y}~ T(y_{C},\tilde{y}\!+\!y_{C},y)~e^{-\pa y_{C}}
 T(y_{A},y_{B},\tilde{y})~=~T(y_{A}\!+\!y_{C},y_{B}\!+\!y_{C},y)~.
\end{array}
\label{balg}
\eeq
The last equation corresponds to the quadratic Leibnitz condition 
\eq{qualeibcond}. 
An important version of the Leibnitz rule reads:
\beq
\int \! d^{d}\tilde{y}~
e^{\pa y_{B}}T(y_{A},\tilde{y}\!+\!y_{B},y\!+\!y_{B})~
e^{\pa y_{C}} T(y_{C},y_{D},\tilde{y}\!+\!y_{C})
~=~(A\leftrightarrow D, B\leftrightarrow C)~.
\label{leibforjac}
\eeq 
This identity was (in the constant case) written down in \cite[Equation
(6.1)]{bering} in the context of the boundary Poisson bracket as a
sufficient condition for the Jacobi identity to hold. A proof is given
in Appendix~\ref{appleibforjac}.

\subsection{Change of Product}
\label{secnonultraadj}

\noi
To deal with the non-ultra-local product, we invoke the following
Ansatz: There exists an invertible matrix-valued smooth function
\mb{\Lambda^{k}{}_{a}} and a fiducial solution 
\mb{c^{a,b}_{(0)c}} such that
\beq
  c^{k,\ell}_{n}~=~\Lambda^{k}{}_{a}~\Lambda^{\ell}{}_{b}~
 c^{a, b}_{(0)c}~ \Lambda^{c}{}_{n}~.
\label{adjexactansatz}
\eeq
It would be obvious to use a ultra-local solution 
\mb{c^{a,b}_{(0)c}~=~ c^{a,b}_{(0)}~\delta^{a+b}_{c}}
as the fiducial solution, but any solution will do. 
We may view \mb{\Lambda^{a}{}_{k}} as a change of 
the renormalization scheme; also known as finite renormalization.

\vs
\noi
We sum up the \mb{\Lambda^{k}{}_{a}} as
\beq
   \Lambda(y,q^{(0)})~=~\sum_{k,a} \frac{y^{k}}{k!}~
\Lambda^{k}{}_{a}~ q^{(0)a}~.
\eeq
The Fourier transform reads
\beq
   \Lambda(y,y^{(0)})~=~
\int\! d^{d}q^{(0)} ~e^{-q^{(0)}y^{(0)}}~ \Lambda(y,q^{(0)}) ~.
\eeq
The Ansatz \eq{adjexactansatz} then reads:
\beq
T(y_{A},y_{B},y) ~=~
\int\! d^{d}y^{(0)}_{A}~ d^{d}y^{(0)}_{B}~ d^{d}y^{(0)}~ 
\Lambda(y_{A},y^{(0)}_{A})~ \Lambda(y_{B},y^{(0)}_{B})~
T^{(0)}(y^{(0)}_{A},y^{(0)}_{B},y^{(0)})~
\Lambda^{-1}(y^{(0)},y)~.
\eeq

\vs
\noi
The Ansatz \eq{adjexactansatz} automatically satisfies the 
commutativity and the associativity requirements. 
The quadratic Leibnitz condition \eq{qualeibcond} is met 
by the following generic condition:
\bea
   \pa_{i}\Lambda^{k}{}_{a}&=& 
\Lambda^{k}{}_{a-e_{i}}- \Lambda^{k+e_{i}}{}_{a}
+\Lambda^{k}{}_{b}~ c^{b,d}_{(0)a} ~ \Lambda^{e_{i}}{}_{d}
-\Lambda^{k}{}_{b}~ c^{b,e_{i}}_{(0)a}~,
\label{lambdaeqadj}
\eea
or equivalently, in terms of the inverse matrix:
\beq
   \pa_{i}\Lambda^{a}{}_{k}~=~ 
- \Lambda^{a}{}_{\ell}~\Lambda^{b}{}_{k}~\pa_{i}\Lambda^{\ell}{}_{b}~=~ 
\Lambda^{a}{}_{k-e_{i}}-\Lambda^{a+e_{i}}{}_{k} 
- \Lambda^{e_{i}}{}_{d}~ c^{d,a}_{(0)b} ~\Lambda^{b}{}_{k}
+ c^{e_{i},a}_{(0)b}~\Lambda^{b}{}_{k}~.
\eeq
This system is integrable.
For an infinitesimal variation, 
\mb{\Lambda^{a}{}_{k}~=~\delta^{a}_{k}+\delta\Lambda^{a}{}_{k}},
the condition \eq{lambdaeqadj} becomes a linearized, homogeneous,
first order partial differential equation:
\beq
\pa_{i}\delta \Lambda^{k}{}_{a}
~=~(A_{i}\delta \Lambda)^{k}{}_{a}~\equiv~
\delta \Lambda^{k}{}_{a-e_{i}}-\delta \Lambda^{k+e_{i}}{}_{a}
+\delta \Lambda^{e_{i}}{}_{d}~c^{d,k}_{(0)a} ~,~~~~~~~~~~~~
[A_{i}, A_{j}]~=~0~,
\label{dlambdaeqadj}
\eeq
with solution
\beq
  \delta \Lambda(x)~=~
\exp\left(\sum_{i=1}^{d}\int_{x^{(0)}}^{x}\! dx^{i}~ A_{i}\right) 
\delta \Lambda(x^{(0)})~, 
\label{lambdasol}
\eeq
independent of the integration path. Other requirements are:
\beq
\begin{array}{rlcl}
\bullet& {\rm Naturalness}&:& \Lambda^{k}{}_{a=0}~=~\delta^{k}_{0}~.\cr\cr
\bullet& {\rm Decoupling}&:&  \Lambda^{k=0}{}_{a}~=~\delta^{0}_{a}~.
\end{array}
\eeq
It is easy to see that the solution \eq{lambdasol} for
\mb{\Lambda(x)~=~{\bf 1}+\delta\Lambda(x)} 
respects naturalness (and decoupling) requirements for all points \mb{x}, 
if it does so for one point \mb{x^{(0)}}.

\setcounter{equation}{0}
\section{Boundary Poisson Bracket}
\label{secbpb}

\noi
A compelling smooth-to-smooth version of the $x$-pointwise boundary 
Poisson bracket \eq{bpb0} is, cf.\ Section~\ref{secintdist}:
\bea
\{f,g\}_{\infty}
&=&C^{\infty} \left[\dedetwo{\left[
\int_{\Sigma}f\right]}{\phi^{A}}
~\omega^{AB}~\dedetwo{\left[
\int_{\Sigma}q\right]}{\phi^{B}} \right]\cr\cr
&=&C^{\infty} \left[\left(\sum_{k}E_{A(k)}f 
\cdot (\pa^{\dagger})^{k}1\right) 
\omega^{AB} \left(\sum_{\ell}E_{B(\ell)}g \cdot
 (\pa^{\dagger})^{\ell}1\right) \right] \cr\cr
&=& C^{\infty} \sum_{k,\ell,n}~ 
E_{A(k)}f~\omega^{AB}~ E_{B(\ell)}g~
c^{k,\ell}_{n}~(\pa^{\dagger})^{n}1 \cr\cr
&=& \sum_{k,\ell,n}~\pa^{n}\left( 
E_{A(k)}f~c^{k,\ell}_{n}~\omega^{AB}~ 
E_{B(\ell)}g~\right)~,~~~~~~~f,g ~\in~ {\cal E}~.
\label{bpbinfty1}
\eea
One of the main points is that the boundary Poisson bracket can be
defined via the last equality as a composition between smooth
functions into smooth functions without ever introducing the boundary 
distributions. Historically 
\cite{soloviev,soloviev2,beringput,solocomp,bering}, this is what 
have been done. But as the first three lines clearly indicate,
the distribution product is ``lurking just beneath the surface''.

\subsection{Mono-Local Tensors}

\noi
The notion of a tensor field acquires some new features when a boundary 
\mb{\pa \Sigma} is present. These are worthwhile pointing out. First
of all, the Poisson bi-vector \mb{\omega^{AB}~=~\omega^{AB}(x,\phi(x))}
is a function of the field variables \mb{\phi^{A}(x)} and
it may have explicit \mb{x}-dependence, but it {\em does not depend on 
the derivatives} \mb{\pa^{k}\phi^{A}(x)}, \mb{k\neq 0}, 
of the field variables. It is antisymmetric \mb{\omega^{BA}=-\omega^{AB}}.
Furthermore, it satisfies the Jacobi identity:
\beq
0~=~ \sum_{{\rm cycl.} A,B,C} \omega^{AD}~P_{D(0)}\omega^{BC}~.
\label{omjac}
\eeq
This implies the Jacobi identity 
\beq
  0~=~ \sum_{{\rm cycl.} f,g,h} \{f,\{g,h\}_{\infty}\}_{\infty}
~,~~~~~~~~~~~~~~f,g,h~\in~ {\cal E}~,
\eeq
for the boundary Poisson bracket \eq{bpbinfty1} itself, as we shall see
in Appendix~\ref{appjac}.

\vs
\noi
The Poisson bi-vector  \mb{\omega^{AB}(x)} transforms covariantly 
as a {\em tensor}, \ie  
\beq
\omega^{AB}~~~ \longrightarrow~~~~
\omega'^{A'B'}~=~P_{A(0)}\phi'^{A'}~\omega^{AB}~P_{B(0)}\phi'^{B'}
\eeq
under a change of coordinates
\beq
\phi^{A}(x)~~~ \longrightarrow~~~~\phi'^{A'}(x)~=~\phi'^{A'}(\phi(x),x)~,
\eeq
which {\em does not involve the derivatives} \mb{\pa^{k}\phi^{A}(x)},
\mb{k\neq 0},  
of the field variables. The presence of the boundary \mb{\pa \Sigma} 
implies that the $2$-tensor \mb{\omega^{AB}(x)} cannot maintain its
form under a more general field transformation. 
Hence, it is natural to  consider only a Poisson bi-vector 
\mb{\omega^{AB}(x)} which does not depend on the derivatives 
\mb{\pa^{k}\phi^{A}(x)}, \mb{k\neq 0}, in the first place 
since the allowed class of field redefinitions cannot induce them later-on.
We should perhaps stress that we continue to allow for
derivatives \mb{\pa^{k}\phi^{A}(x)} inside \mb{f} and \mb{g} in 
\eq{bpb0}. After all, this is the raison d'etre of a boundary 
Poisson bracket, while on the other hand, the applications usually deal
with a Poisson structure on Darboux form, \ie 
where \mb{\omega^{AB}} is a constant.

\subsection{Multi-Local Formalism}

\noi
Let us mention, without going into lengthy details, 
that to allow for arbitrary field redefinitions, 
one should consider a bi-local Poisson bracket
\bea
\{F,G\}&=&\int_{\Sigma\times\Sigma } \rho~d^{d}x_{A}~
\rho~d^{d}x_{B}~\dedetwo{F}{\phi^{A}(x_{A})}
~\omega^{AB}(x_{A},x_{B})~ \dedetwo{G}{\phi^{B}(x_{B})} \cr\cr\cr
0&=& \sum_{{\rm cycl.} A,B,C}\int_{\Sigma } \rho~d^{d}x_{D}~ 
\omega^{AD}(x_{A},x_{D})~
\dedetwo{\omega^{BC}(x_{B},x_{C})}{\phi^{D}(x_{D})}~.
\label{biomjac}
\eea
By the word {\em bi-local} we are merely referring to that
\mb{\omega^{AB}(x_{A},x_{B})} (in the distributional sense) may depend
on \mb{x_{A}} and \mb{x_{B}}. It may very well be that the support is 
along the diagonal \mb{x_{A}=x_{B}}, but this does not necessarily 
imply that the Poisson bi-vector \mb{\omega^{AB}(x_{A},x_{B})}
can be written with one entry only, due to the very nature of 
distributions.  Whenever one can bring \mb{\omega^{AB}(x_{A},x_{B})} on
the diagonal form 
\beq
\omega^{AB}(x_{A},x_{B})~=~\omega^{AB}(x_{A})~ 
\delta_{\Sigma}(x_{A},x_{B})~,
\label{diagansatz}
\eeq
one may reduce the boundary Poisson bracket \eq{biomjac} 
to the mono-local case.
The point is twofold:
\begin{itemize}
\item
The diagonal Ansatz \eq{diagansatz} is not stabile under general 
field redefinitions, when a boundary \mb{\pa\Sigma} is present.  
\item
Even if  \mb{\omega^{AB}(x_{A},x_{B})} is of
the diagonal form \eq{diagansatz},
the pertinent Jacobi condition \eq{biomjac} may not acquire a
mono-local form, if \mb{\omega^{AB}(x_{A})} depends on the field
derivatives \mb{\pa^{k}\phi^{A}(x)}, \mb{k\neq 0}.
\end{itemize}

\vs
\noi
Here the Dirac delta distribution is defined as
\beq
\forall f~:~~~~~
  \int_{\Sigma\times\Sigma} \rho~d^{d}x_{A}~\rho~d^{d}x_{B}~
f(x_{A},x_{B})~\delta_{\Sigma}(x_{A},x_{B})
~=~\int_{\Sigma } \rho~d^{d}x~f(x,x)~.
\eeq
It has the property that
\beq
\left(\pa^{\dagger}_{x^{i}_{A}}-\pa_{x^{i}_{B}}\right)
\delta_{\Sigma}(x_{A},x_{B})~=~0~.
\eeq

\vs
\noi
The functional derivative may formally be written as
\beq
  \dedetwo{}{\phi^{A}(x)}~=~ 
\sum_{i}\sum_{k}(\pa_{x}^{\dagger})^{k}\delta_{\Sigma}(x,z_{(i)})~ 
P^{(z_{(i)})}_{A(k)}~,
\eeq
where the \mb{i}-sum extends over the finitely many $d$-tuple variables 
\mb{z_{(i)}=(z^{1}_{(i)}, \ldots, z^{d}_{(i)})} of the implicitly 
attacked multi-local functional. That is, the functional derivative 
operation ``peers'' beyond the integral symbols of the unwritten 
multi-local functional (if any integrals).

\vs
\noi
In the multi-local framework, the proof of the Jacobi identity for the 
boundary Poisson bracket will follow from the 
traditional argument: Namely, as a consequence of 1) the Jacobi identity 
\eq{biomjac} for \mb{\omega^{AB}(x_{A},x_{B})}, 2) the Leibnitz rule 
and 3) the fact that two functional derivatives commute:
\beq
 \left[ ~\dedetwo{}{\phi^{A}(x_{A})}~,~
\dedetwo{}{\phi^{B}(x_{B})}~ \right]~=~0~.
\eeq
While multi-local objects are a straightforward
and natural generalization, we shall devote our attention
to the important mono-local case. First of all, because it is
interesting in its own right. Secondly, since we have only sketched 
the multi-local formalism, we would like to provide a proof of
the Jacobi identity that rest within the mono-local formalism itself. 
(See Appendix~\ref{appjac}). Unfortunately, the mono-local proof
does not have the attractive simplicity of the multi-local proof.

\subsection{Fourier Transform}

\noi
It is convenient to resum the higher derivatives in a series,
\bea
  P_{A}(q)f&\equiv&\sum_{k=0}^{\infty} q^{k}~P_{A(k)}f~, \cr
E_{A}(q)f& \equiv &\sum_{k=0}^{\infty} q^{k}~E_{A(k)}f~
~=~\exp\left[-\pa \papa{q}\right]P_{A}(q)f~,
\label{peq}
\eea
and to introduce the Fourier transform
\bea
P_{A}(y)f&\equiv&\int d^{d}q~e^{-qy} P_{A}(q)f~, \cr
E_{A}(y)f&\equiv&\int d^{d}q~e^{-qy} E_{A}(q)f 
~=~e^{-\pa y} P_{A}(y)f~.
\label{pey}
\eea
Then we may write a functional derivative as
\beq
 \dedetwo{\left[\int_{\Sigma}f\right]}{\phi^{A}}
~=~\int \! d^{d}y~e^{\pa^{\dagger}y} P_{A}(y)f
~=~\int \! d^{d}y~ E_{A}(y)f \cdot e^{\pa^{\dagger}y}1 ~.
\label{localdiff3}
\eeq
As a consequences of \eq{heisalg}, we have the following Heisenberg 
group structure \cite{beringput}:
\bea
 e^{\pa y} P_{A}(y_{A})&=&P_{A}(y_{A}\!+\!y) e^{\pa y}~,\cr\cr
 E_{A}(y_{A})&=&E_{A}(y_{A}\!+\!y) e^{\pa y}~,\cr\cr
P_{A}(y_{A}) e^{\pa^{\dagger} y}
&=&e^{\pa^{\dagger} y} P_{A}(y_{A}\!+\!y)~,\cr\cr
E_{A}(y_{A}) e^{\pa^{\dagger} y} &=&E_{A}(y_{A}\!+\!y)~.
\label{heisgroup}
\eea 
The $x$-pointwise boundary Poisson bracket \eq{bpbinfty1} can be
stated as
\bea
\{f,g\}_{\infty}
&=&C^{\infty} \left[\dedetwo{\left[\int_{\Sigma}f\right]}{\phi^{A}}
~\omega^{AB}~\dedetwo{\left[\int_{\Sigma}q\right]}{\phi^{B}} \right]\cr\cr
&=&C^{\infty}\int \!  d^{d}y_{A}~d^{d}y_{B}~d^{d}y~
E_{A}(y_{A})f~\omega^{AB}~ E_{B}(y_{B})g~
T(y_{A},y_{B},y)~ e^{\pa^{\dagger}y}1 \cr\cr
&=& \int \!  d^{d}y_{A}~d^{d}y_{B}~d^{d}y~ e^{\pa y}\left(
E_{A}(y_{A})f~\omega^{AB}~T(y_{A},y_{B},y)~E_{B}(y_{B})g\right)~.
\label{bpbinfty2}
\eea
The Jacobi identity for this un-integrated, $x$-pointwise boundary 
Poisson bracket \mb{\{f,g\}_{\infty}} is satisfied. Here (and 
everywhere else in this paper), we are referring to the Jacobi identity
in its strongest form, \ie without any residual total derivative 
terms nor any total adjoint derivative terms leftover. It is proven 
in Appendix~\ref{appjac}. Interestingly, it continues to hold if we 
replace the smooth functions \mb{f}, \mb{g} and \mb{\omega^{AB}} 
with general boundary distributions \mb{u}, \mb{v} and \mb{\omega^{AB}} 
and take the last equation in \eq{bpbinfty1} or in \eq{bpbinfty2} 
as a definition:
\beq
  0~=~ \sum_{{\rm cycl.} u,v,w} \{u,\{v,w\}_{\infty}\}_{\infty} 
  ~,~~~~~~~~~~~~~~ u,v,w ~\in~ {\cal B}~.
\label{uvwjac}
\eeq

\setcounter{equation}{0}
\section{Alternative $\Delta^{(n)}1$ Basis}
\label{secalt}

\subsection{Heuristic Remarks}
\label{secheur}

\noi
Heuristically, the adjoint basis \mb{(\pa^{\dagger})^{k}1} may be
viewed as a renormalization/definition of what should be meant by
the singular expression 
\mb{(\rho 1_{\Sigma})^{-1}(-\pa)^{k}(\rho 1_{\Sigma})}, where
\beq
 1_{\Sigma}(x)~\equiv~\left\{\begin{array}{rl} 1 & 
{\rm if}~ x \in \Sigma  \\
 0 & {\rm otherwise} \end{array} \right.
\eeq
is the characteristic function for the set \mb{\Sigma} inside \mb{\R^{d}}.

\vs
\noi
In the same way, we shall now consider an alternative basis 
\mb{\Delta^{(n)}1}, which provides a rigorous meaning to 
\mb{\pa^{n} \ln(\rho 1_{\Sigma})}. It will be convenient in the
rigorous treatment to exclude \mb{n=0}. 

\vs
\noi
For the ultra-local product, we shall use this basis to argue that
the parameter \mb{s} itself has an interpretation as a renormalization of
\mb{\ln(\rho 1_{\Sigma})/\Gamma(0)}.

\subsection{$\Delta^{(n)}$ Operators}
\label{secdeltaop}

\noi
After the above heuristic introduction, we turn to a rigorous
definition of the operators 
\mb{\Delta^{(n)}:{\cal B}\to {\cal B}} for 
\mb{n\in \N_{0}^{d} \backslash \{0\}}:
\bea
\Delta(y)u&\equiv& \sum_{n \neq 0} \frac{y^{n}}{n!}~ \Delta^{(n)}u
~\equiv~\!\! \int_{0}^{1} \frac{d \alpha}{e^{-\pa^{\dagger}\alpha y}1}~
\frac{d}{d \alpha} e^{-\pa^{\dagger}\alpha y}u\cr\cr
&=&\!\! \int_{0}^{1} \frac{d \alpha}{\alpha}~
\frac{1}{e^{-\pa^{\dagger}\alpha y}1} ~
y\frac{\pa}{\pa y} e^{-\pa^{\dagger}\alpha y}u
 ~,~~~~~~~~~u \in {\cal B}~,\cr\cr
\Delta(y\!=\!0)u&=&0~.
\label{deltadef}
\eea
The integral expressions should be understood by first Taylor
expanding around \mb{y=0} and then performing the 
\mb{\alpha}-integration. Up to second order in \mb{y}, we get:
\bea
   \Delta^{(e_{i})}u&=&-\pa_{i}^{\dagger}u~, \cr\cr
   \Delta^{(e_{i}+e_{j})}u&=&
\frac{1}{2}\left(\pa_{i}^{\dagger}\pa_{j}^{\dagger}u
-\pa_{i}^{\dagger}1 \cdot \pa_{j}^{\dagger}u \right)
~+~ (i \leftrightarrow j)~.
\eea
Note that this definition relies heavily on the existence of the 
distribution product \eq{proddef}. For \mb{u\equiv 1} the definition
\eq{deltadef} simplifies to 
\beq
  \Delta(y)1~=~ \ln\left(e^{-\pa^{\dagger} y}1 \right)~.
\eeq
An equivalent definition is given by
\beq
y\frac{\pa}{\pa y}\Delta(y)u~=~ 
\sum_{n \neq 0} \frac{|n|~y^{n}}{n!}~ \Delta^{(n)}u
~=~\frac{1}{e^{-\pa^{\dagger} y}1} ~
y\frac{\pa}{\pa y} e^{-\pa^{\dagger} y}u~,~~~~~u \in {\cal B}~. 
\eeq

\subsection{Inverse Relations}
\label{secdeltaopinv}

\noi
Of particular interest is the case where the
\mb{\Delta^{(n)}1}, \mb{n\neq 0}, become linearly independent. 
Because of the triangular form of the  \mb{\Delta^{(n)}1}, \mb{n\neq 0}, 
this happens precisely when
\beq
 \forall n \neq 0~:~ \Delta^{(n)}1 \neq 0~.
\eeq
Granted that this is the case, the inverse relations read:
\bea
e^{\pa^{\dagger} y}u&\equiv& \sum_{k}
\frac{y^{k}}{k!}~(\pa^{\dagger})^{k}u
~=~ \int_{0}^{1} \! d \alpha~ \exp\left(\Delta(-\alpha y)1\right) \cdot
\frac{d}{d \alpha}\Delta(-\alpha y)u \cr\cr
&=&\int_{0}^{1} \! \frac{d \alpha}{\alpha}~ 
\exp\left(\Delta(-\alpha y)1\right) \cdot
\frac{\pa}{\pa y}\Delta(-\alpha y)u~,~~~~~u \in {\cal B}~,\cr\cr
\frac{\pa}{\pa y}e^{\pa^{\dagger} y}u&=& \sum_{k}
\frac{|k|~y^{k}}{k!}~(\pa^{\dagger})^{k}u
~=~\exp\left(\Delta(- y)1\right) \cdot
\frac{\pa}{\pa y}\Delta(- y)u~,\cr\cr
e^{\pa^{\dagger} y}1&=&\exp\left(\Delta(-y)1\right)~
\equiv~\sum_{k}y^{k}~
S^{(k)}\left(\{ \frac{(-1)^{n}}{n!}\Delta^{(n)}1 \}_{n \neq 0}\right)~. 
\eea
In the last line we have indicated that the adjoint basis elements
\bea
\frac{(\pa^{\dagger})^{k}1}{k!}&=&
S^{(k)}\left(\{ \frac{(-1)^{n}}{n!} 
\Delta^{(n)}1\}_{n \neq 0}\right) \cr\cr
&=&\left(-1\right)^{k}\!\!\!\!\!\!\!\!\!\!\!\!\!\!\sum_{
\begin{array}{c}
\{m_{n}\}_{n\neq 0} \cr 
k=\sum_{n\neq 0}m_{n}n
\end{array}
}\!\!\!\!\!\!\!\!
 \prod_{n \neq 0}
\frac{1}{m_{n}!}\left( \frac{1}{n!} \Delta^{(n)}1 \right)^{m_{n}}
~,~~~~~~~~~~~~~~~k \in \N_{0}^{d}~,
\eea
up to a normalization, become the \mb{d}-dimensional Schur polynomials 
of the alternative basis elements \mb{\frac{(-1)^{n}}{n!}\Delta^{(n)}1},
\mb{n\in \N_{0}^{d} \backslash \{0\}}.

\subsection{Derivative of $\Delta^{(n)}$ operators}

\noi
The derivative \mb{\pa_{i}} acts very simple on the \mb{\Delta^{(n)}} 
operators, as the reader might have suspected from the heuristic arguments
given in Section~\ref{secheur}:
\bea
 \pa_{i}\Delta^{(n)}u&=&\Delta^{(n+e_{i})}u~, \cr\cr
e^{\pa \bar{y}}\Delta(y)u&=&\Delta(y\!+\!\bar{y})u-\Delta(\bar{y})u~.
\eea  
This follows straightforwardly from the definitions \eq{defder} and 
\eq{deltadef}.
Note that we can describe the derivative \mb{\pa_{i}} in terms of the
\mb{\Delta^{(n)}} operators {\em without} using the distribution
product \eq{proddef}. 
This is the key advantage of the \mb{\Delta^{(n)}} operators.
As one might expect, the action of the adjoint derivatives 
\mb{\pa_{i}^{\dagger}} now becomes complicated:
\beq
e^{\pa^{\dagger} \bar{y}} \Delta(y)u
~=~ \frac{\Delta(y\!-\!\bar{y})u-\Delta(-\bar{y})u}
{\exp\left( \Delta(-\bar{y})1 \right)}~.
\eeq

\subsection{Product in $\Delta^{(n)}1$ Basis}

\noi
Starting from the \mb{\Delta^{(n)}1} basis, we can describe the 
distribution product as
\beq
  \Delta^{(k)}1 \cdot \Delta^{(\ell)}1~=~
 \sum_{n \neq 0} \tilde{c}^{k,\ell}_{n}~\Delta^{(n)}1~, 
\label{proddef2}
\eeq
for some smooth structure functions 
\mb{\tilde{c}^{k,\ell}_{n} \in {\cal E}}.
From this point of view we would have the same basic requirements,
such as the commutativity and the associativity assumption 
and the Leibnitz rule. (The naturalness assumption is automatically 
fulfilled, because \mb{n \neq 0}.)

\vs
\noi
The product \eq{proddef} is summed up in 
\beq
\tilde{T}(y_{A},y_{B},q)~\equiv~\sum_{k,\ell,n \neq 0} 
\frac{y_{A}^{k}}{k!}~\frac{y_{B}^{\ell}}{\ell!} ~
\tilde{c}^{k,\ell}_{n} ~q^{n}~.
\eeq
It has Fourier transform
\beq
\tilde{T}(y_{A},y_{B},y)~\equiv~
\int \! d^{d}q~e^{-q y}~\tilde{T}(y_{A},y_{B},q)~.
\eeq
We note for later that:
\beq
\int \! d^{d}y~\tilde{T}(y_{A},y_{B},y)~=~
\tilde{T}(y_{A},y_{B},q\!=\!0)~=~0~.
\eeq
The closeness \eq{proddef2} of the boundary algebra then reads
\beq
\Delta(y_{A})1 \cdot \Delta(y_{B})1 ~=~
\int d^{d}y~\tilde{T}(y_{A},y_{B},y)~\Delta(y)1~.
\eeq
Contrary to the adjoint basis, the Leibnitz condition is {\em linear} 
in the structure functions
\bea 
 \pa_{i} \tilde{c}^{k,\ell}_{n} &=&\tilde{c}^{k+e_{i},\ell}_{n}
+\tilde{c}^{k,\ell+e_{i}}_{n}-\tilde{c}^{k,\ell}_{n-e_{i}}~, \cr\cr
e^{\pa \bar{y}}\tilde{T}(y_{A},y_{B},y\!-\!\bar{y})
&=& \tilde{T}(y_{A}\!+\!\bar{y},y_{B}\!+\!\bar{y},y)
- \tilde{T}(y_{A}\!+\!\bar{y},\!\bar{y},y) \cr\cr
 &&-\tilde{T}(\bar{y},y_{B}\!+\!\bar{y},y)
+ \tilde{T}(\bar{y},\!\bar{y},y)~.
\label{linleibcond}
\eea

\subsection{Ultra-Local Product}
\label{secultradel}

\noi
We have that 
\mb{\tilde{c}^{k,\ell}_{n}~=~\tilde{c}^{k,\ell}~\delta^{k+\ell}_{n}} 
is ultra-local in the \mb{\Delta^{(n)}1} basis if and only if 
\mb{c^{k,\ell}_{n}~=~c^{k,\ell}~\delta^{k+\ell}_{n}} 
is ultra-local in the adjoint basis \mb{(\pa^{\dagger})^{k}1}. 
This is due to the homogeneity of the \mb{\Delta^{(n)}} 
operators of homogeneity weights \mb{n \in \N_{0}^{d}}.
Here the homogeneity weights are wrt.\ a scaling of the adjoint 
differential operator 
\mb{\pa_{i}^{\dagger} \to \lambda_{i}\pa_{i}^{\dagger}}, 
\mb{i=1,\ldots,d}. Or vice-verse. Only the Leibnitz rule is slightly 
changed compared with the adjoint case in Section~\ref{secultraadj}. 
It still splits into two conditions:
\beq
 \tilde{c}^{k+e_{i},\ell}+ \tilde{c}^{k,\ell+e_{i}} 
~=~ \tilde{c}^{k,\ell}~
~~~~\wedge ~~~~~~\pa_{i} \tilde{c}^{k,\ell}_{n}~=~0~.
\eeq
The generic solutions are \mb{2}-coboundaries of locally constant 
invertible elements:
\beq 
\exists ~\tilde{b}^{k} \in {\cal E}^{\times}~:~
\tilde{c}^{k,\ell}~=~\frac{\tilde{b}^{k}~
\tilde{b}^{\ell}}{\tilde{b}^{k+\ell}}~,
\eeq
that satisfies the first of the Leibnitz conditions -- below written
slightly simplified by use of the exactness property:
\beq
 \frac{1}{\tilde{c}^{k,e_{i}}}+\frac{1}{\tilde{c}^{\ell,e_{i}}}
~=~\frac{1}{\tilde{c}^{k+\ell,e_{i}}}~.
\eeq
It follows that there exists constants 
\mb{s_{ij}=s_{ji}}, \mb{i,j=1,\ldots,d}, such that 
\beq
 \frac{1}{\tilde{c}^{k,e_{i}}}~=~\sum_{j=1}^{d}\frac{k_{j}}{s_{ij}}~.
\eeq
A check of the product confirms that these constants \mb{s_{ij}}
coincide with the \mb{s_{ij}} introduced in the adjoint case 
in Section~\ref{secultraana}.
In fact, the ultra local solution \eq{mainresult} becomes
\bea
  \tilde{c}^{k,\ell}&=&\tilde{c}_{(I)}^{k,\ell}~
\tilde{c}_{(II)}^{k,\ell}~, \cr\cr
 \tilde{c}_{(I)}^{k,\ell}&=&s~ B(i(k),i(\ell)) ~, \cr\cr
\tilde{c}_{(II)}^{k,\ell}&=&\prod_{\begin{array}{c}
 i,j \cr i \neq j \cr s_{ij}=0\end{array}} 
\delta_{0}^{{\rm min}(k_{i}+\ell_{i},k_{j}+\ell_{j})}~,
\label{mainresult2}
\eea
\ie all the diagonal parameter \mb{s} enter multiplicatively. 
The arguments \mb{i(k)} in the Euler Beta function are 
non-negative integers. Recall that the Euler Beta function has
a pole located in zero.  All other branches besides
the $1$-dimensional branch with the horizontal Young diagram 
\mb{(\young_{1}\!=\!d)} are ``infected'' with this pole.
This is an indirect manifestation of the fact that the generators 
\mb{\Delta^{(n)}1}, \mb{n\neq 0}, on the other branches are not 
linearly independent.

\subsection{Non-Ultra-Local Product}
\label{secnonultradel}

\noi
To deal with the non-ultra-local case, we invoke the following
Ansatz: There exists an invertible matrix-valued smooth function
\mb{\Lambda^{k}{}_{a}} and a fiducial solution 
\mb{\tilde{c}^{a,b}_{(0)c}} such that
\beq
  \tilde{c}^{k,\ell}_{n}~=~\sum_{a,b,c\neq 0}
\Lambda^{k}{}_{a}~\Lambda^{\ell}{}_{b}~
 \tilde{c}^{a, b}_{(0)c}~ \Lambda^{c}{}_{n}~.
\label{delexactansatz}
\eeq
It would be obvious to use the ultra-local solution 
\mb{\tilde{c}^{a,b}_{(0)c}~=~B(|a|,|b|)~\delta^{a+b}_{c}} 
as the fiducial solution, but in fact any solution will do. 

\vs
\noi
Again we sum up the \mb{\Lambda^{k}{}_{a}} as
\beq
   \Lambda(y,q^{(0)})~=~\sum_{k,a\neq 0} \frac{y^{k}}{k!}~
\Lambda^{k}{}_{a}~ q^{(0)a}~,
\eeq
and the Fourier transform
\beq
   \Lambda(y,y^{(0)})~=~
\int\! d^{d}q^{(0)} ~e^{-q^{(0)}y^{(0)}}~ \Lambda(y,q^{(0)}) ~.
\eeq
We note for later that:
\beq
\int \! d^{d}y^{(0)}~\Lambda(y,y^{(0)})~=~
\Lambda(y,q^{(0)}\!=\!0)~=~0~.
\eeq
The Ansatz \eq{delexactansatz} then reads:
\beq
\tilde{T}(y_{A},y_{B},y) ~=~
\int\! d^{d}y^{(0)}_{A}~ d^{d}y^{(0)}_{B}~ d^{d}y^{(0)}~ 
\Lambda(y_{A},y^{(0)}_{A})~ \Lambda(y_{B},y^{(0)}_{B})~
\tilde{T}^{(0)}(y^{(0)}_{A},y^{(0)}_{B},y^{(0)})~
\Lambda^{-1}(y^{(0)},y)~.
\eeq

\vs
\noi
The Ansatz \eq{delexactansatz} automatically satisfies the 
commutativity and the associativity requirements. The linear Leibnitz 
condition \eq{linleibcond} is met by the following generic condition:
\bea
  \pa_{i}\Lambda^{k}{}_{a}&=&
\Lambda^{k+e_{i}}{}_{a}-\Lambda^{k}{}_{a-e_{i}}~, \cr\cr
e^{(q^{(0)}+\pa) \bar{y}}~\Lambda(y,q^{(0)})
&=&\Lambda(y\!+\!\bar{y},q^{(0)})-\Lambda(\bar{y},q^{(0)})~, \cr\cr
e^{\pa \bar{y}}~\Lambda(y,y^{(0)}\!-\!\bar{y})
&=&\Lambda(y\!+\!\bar{y},y^{(0)})-\Lambda(\bar{y},y^{(0)})~.
\label{lambdaeqdel}
\eea
or equivalently, in terms of the inverse matrix:
\beq
\pa_{i}\Lambda^{a}{}_{k}~=~ 
 -\Lambda^{a}{}_{\ell}~\Lambda^{b}{}_{k}~\pa_{i}\Lambda^{\ell}{}_{b}
~=~ \Lambda^{a+e_{i}}{}_{k}-\Lambda^{a}{}_{k-e_{i}}~.
\eeq
The shift operators  
\mb{(E^{n})^{k}{}_{\ell}~=~\delta^{k+n}_{\ell}} span a commutative and
associative algebra 
\mb{\dbar{E}^{n}\dbar{E}^{m}=\dbar{E}^{n+m}}, with a trivial 
differental structure \mb{[\pa_{i},\dbar{E}^{n}]~=~0}.
Hence, we can write the condition \eq{lambdaeqdel} as
\beq
 [\pa_{i}-\dbar{E}^{e_{i}}, \dbar{\Lambda}]~=~0~.
\eeq
This system is clearly integrable with solution
\bea
\dbar{\Lambda}(x\!+\!\bar{y})&=&
\exp\left(\sum_{i=1}^{d}\bar{y}^{i}~\dbar{E}^{e_{i}}\right)~
\dbar{\Lambda}(x)~
\exp\left(-\sum_{i=1}^{d}\bar{y}^{i}~\dbar{E}^{e_{i}}\right) \cr\cr
&=& \sum_{k,a} \frac{\bar{y}^{k}}{k!}~\dbar{E}^{k} ~\dbar{\Lambda}(x)~ 
\frac{(-\bar{y})^{a}}{a!}~\dbar{E}^{a}~, \cr\cr 
\Lambda(x\!+\!\bar{y})^{k}{}_{a}&=& 
\sum_{\ell,b} \frac{\bar{y}^{\ell}}{\ell!} 
~\Lambda(x)^{k+\ell}{}_{a-b}~ 
\frac{(-\bar{y})^{b}}{b!}~,
\eea
where \mb{x} is a reference point. This can of course also be
derived by expanding the second expression in \eq{lambdaeqdel}.

\setcounter{equation}{0}
\section{Conclusions}
\label{conclus}

\noi
When imposing the Jacobi identity for a boundary 
Poisson bracket, an infinite tower of conditions emerges. 
(Plus an expected standard Jacobi condition on the Poisson bi-vector
\mb{\omega^{AB}}.)
In the ultra-local case, this tower was already known \cite{bering}.
We have demonstrated that there is a well-defined distribution product 
behind the boundary Poisson bracket construction. 
We have shown that the tower of conditions on the boundary Poisson bracket
can really be viewed as conditions on this distribution product instead. 
As a result, we have simplified the boundary calculus of  Hamiltonian field 
theory considerably. We hope that these developments will stimulate
further interests in the physical systems with a boundary.


\vs
\noi
{\bf Acknowledgements}.
We would like to thank B.~Morariu for carefully reading the manuscript.
The work is supported in part by U.S.\ Department of Energy (DoE)
under grant no.~DE-FG02-91ER-40651-Task-B.

\begin{appendix}

\setcounter{equation}{0}
\section{Proof of the Leibnitz Identity \eq{leibforjac}}
\label{appleibforjac}

\noi
Consider the following \mb{(A\leftrightarrow D, B\leftrightarrow C)}
symmetric object:
\bea
\lefteqn{e^{\pa y_{B}}e^{\pa^{\dagger} y_{A}}1
\cdot e^{\pa y_{C}}e^{\pa^{\dagger} y_{D}}1}\cr\cr 
&=&e^{-\pa^{\dagger} y_{C}}\left[ 
e^{\pa (y_{B}-y_{C})}e^{\pa^{\dagger} y_{A}}1
\cdot e^{\pa^{\dagger} y_{C}}1 
\cdot e^{\pa^{\dagger} y_{D}}1 \right]  \cr\cr
&=&e^{-\pa^{\dagger} y_{C}}\left[ 
e^{\pa (y_{B}-y_{C})}e^{\pa^{\dagger} y_{A}}1
\cdot\int \! d^{d}\tilde{y}~ T(y_{C},y_{D},\tilde{y}\!+\!y_{C})~
e^{\pa^{\dagger} (\tilde{y}+y_{C})}1 \right] \cr\cr
&=&\int \! d^{d}\tilde{y}~e^{-\pa^{\dagger} y_{B}}\left[ 
e^{\pa^{\dagger} y_{A}}1
\cdot e^{\pa (y_{C}-y_{B})}T(y_{C},y_{D},\tilde{y}\!+\!y_{C})~
e^{\pa^{\dagger} (\tilde{y}+y_{B})}1 \right] \cr\cr
&=&\int \! d^{d}\tilde{y}~e^{-\pa^{\dagger} y_{B}}\left[ 
\int \! d^{d}y~T(y_{A},\tilde{y}\!+\!y_{B},y\!+\!y_{B})~ 
e^{\pa^{\dagger} (y+y_{B})}1~
e^{\pa (y_{C}-y_{B})}T(y_{C},y_{D},\tilde{y}\!+\!y_{C})~
 \right] \cr\cr
&=&\int \! d^{d}y~d^{d}\tilde{y}~
e^{\pa y_{B}}T(y_{A},\tilde{y}\!+\!y_{B},y\!+\!y_{B})~
e^{\pa y_{C}} T(y_{C},y_{D},\tilde{y}\!+\!y_{C})~e^{\pa^{\dagger} y}1 ~.
\label{leibforjac2}
\eea 
Here we have applied the Leibnitz rule three times plus various definitions.
The desired Leibnitz identity \eq{leibforjac} now follows from 
the unique decomposition of a boundary distribution into its
\mb{C^{\infty}} coefficient functions, cf.\ \eq{boundistdecomp}.
\proofbox

\setcounter{equation}{0}
\section{Mono-Local Proof of the Jacobi Identity \eq{uvwjac}}
\label{appjac}

\noi
Consider three boundary distributions \mb{u},\mb{v} and \mb{w}. 
When calculating a ``double bracket'', using
the last formula in \eq{bpbinfty2}, we get three terms:
\beq
 \{ u, \{v, w \}_{\infty} \}_{\infty}
~=~ T_{1}(u,v,w)+T_{2}(u,v,w)-T_{1}(u,w,v)~.
\label{doubletrouble}
\eeq
Here the first term \mb{T_{1}} is
\bea
T_{1}(u,v,w)&\equiv& \! \! \! \!  \int d^{6d}y~e^{\pa y}\left[ 
E_{A}(y_{A})u~T(y_{A},y_{B},y)~\omega^{AB}~e^{-\pa y_{B}}P_{B}(y_{B})
e^{\pa (\tilde{y}-y_{C})} P_{C}(y_{C})v \right. \cr\cr
&&\left. \times~e^{\pa (\tilde{y}-y_{B})}\left( 
\omega^{CD}~T(y_{C},y_{D},\tilde{y})~E_{D}(y_{D})w \right)\right]\cr\cr
&\equi{\eq{heisgroup}}& \! \! \! \!  \int d^{6d}y~e^{\pa y}\left[ 
E_{A}(y_{A})u~T(y_{A},y_{B},y)~\omega^{AB}~
e^{\pa (\tilde{y}- y_{B}- y_{C})}
P_{B}(y_{B}\!+\!y_{C}\!-\!\tilde{y}) P_{C}(y_{C})v \right. \cr\cr
&&\left. \times~e^{\pa (\tilde{y}-y_{B})}\left( 
\omega^{CD}~T(y_{C},y_{D},\tilde{y})~E_{D}(y_{D})w \right)\right]\cr\cr
&=& \! \! \! \!  \int d^{6d}y~e^{\pa y'}\left[ 
e^{\pa  y'_{B}}\left(E_{A}(y_{A})u~
T(y_{A},\tilde{y}'\!+\!y'_{B},y\!+\!y'_{B})~\omega^{AB}\right)
P_{B}(y'_{B}) P_{C}(y_{C})v \right. \cr\cr
&&\left. \times~e^{\pa y_{C}}\left( 
\omega^{CD}~T(y_{C},y_{D},\tilde{y}'\!+\!y_{C})~E_{D}(y_{D})w
\right)\right]\cr\cr
&\equi{\eq{leibforjac}}& T_{1}(w,v,u)~.
\eea
We have used the following shorthand notation for the integration measure
\beq
 d^{6d}y~\equiv~d^{d}y_{A}~ d^{d}y_{B}~d^{d}y~ 
d^{d}y_{C}~ d^{d}y_{D}~d^{d}\tilde{y}~,
\eeq
and we have performed the following change of integration variables
\beq
\sixbyone{y'_{A}}{y'_{B}}{y'}{y'_{C}}{y'_{D}}{\tilde{y}'}~=~
\left( \begin{array}{ccccccccccc}
{1}&&{0}&&{0}&&{0}&&{0}&&{0}\cr{0}&&{1}&&{0}&&{1}&&{0}&&{-1}\cr
{0}&&{-1}&&{1}&&{-1}&&{0}&&{1}\cr{0}&&{0}&&{0}&&{1}&&{0}&&{0}\cr
{0}&&{0}&&{0}&&{0}&&{1}&&{0}\cr{0}&&{0}&&{0}&&{-1}&&{0}&&{1} 
\end{array} \right)
\sixbyone{y_{A}}{y_{B}}{y}{y_{C}}{y_{D}}{\tilde{y}}~,
\eeq
which has the Jacobian equal to $1$. The full Jacobi identity 
contains six \mb{T_{1}} terms. 
It is easy to see that they cancel by symmetry.

\vs
\noi
The second term \mb{T_{2}} in \eq{doubletrouble} is
\bea
T_{2}(u,v,w)&\equiv& \! \! \! \!  \int d^{6d}y~e^{\pa y}\left[ 
E_{A}(y_{A})u~T(y_{A},y_{B},y)~\omega^{AB}~e^{-\pa y_{B}}P_{B}(y_{B})
e^{\pa \tilde{y}}\omega^{CD} \right. \cr\cr
&&\left. \times~e^{\pa (\tilde{y}-y_{B})}\left(  E_{C}(y_{C})v
~T(y_{C},y_{D},\tilde{y})~E_{D}(y_{D})w \right)\right]\cr\cr
&\equi{\eq{heisgroup}}& \! \! \! \!  \int d^{6d}y~e^{\pa y}\left[ 
E_{A}(y_{A})u~T(y_{A},y_{B},y)~\omega^{AB}~ \right. \cr\cr
&&\left. \times~e^{\pa (\tilde{y}-y_{B})}
\left(P_{B}(y_{B}\!-\!\tilde{y})\omega^{CD}~E_{C}(y_{C})v
~T(y_{C},y_{D},\tilde{y})~E_{D}(y_{D})w \right)\right]\cr\cr
&=& \! \! \! \!  \int d^{6d}y~e^{\pa y}\left[ 
E_{A}(y_{A})u~T(y_{A},y_{B}'\!+\!\tilde{y},y)~\omega^{AB}~ \right. \cr\cr
&&\left. \times~e^{-\pa y'_{B}}
\left(P_{B}(y'_{B})\omega^{CD}~E_{C}(y_{C})v
~T(y_{C},y_{D},\tilde{y})~E_{D}(y_{D})w \right)\right]\cr\cr
&=& \! \! \! \!  \int d^{6d}y~e^{\pa y}\left[ 
E_{A}(y_{A})u~T(y_{A},\tilde{y}\!+\!y_{B}',y)~\omega^{AB}~ \right. \cr\cr
&&\left. \times~e^{-\pa y'_{B}}
\left(\delta^{d}(y'_{B})~P_{B(0)}\omega^{CD}~E_{C}(y_{C})v
~T(y_{C},y_{D},\tilde{y})~E_{D}(y_{D})w \right)\right]\cr\cr
&=& \! \! \! \!  \int d^{5d}y~e^{\pa y}\left[ 
E_{A}(y_{A})u~T(y_{A},\tilde{y},y)~\omega^{AB}~ \right. \cr\cr
&&\left. \times~
P_{B(0)}\omega^{CD}~E_{C}(y_{C})v
~T(y_{C},y_{D},\tilde{y})~E_{D}(y_{D})w \right]~.
\eea
In the next-to-last step we used the assumption that 
\mb{\omega^{CD}} does not depend on the derivatives
\mb{\pa^{k}\phi^{A}(x)}, \mb{k\neq 0}. It is now easy to see that 
in the full Jacobi identity, the three \mb{T_{2}} terms cancel by 
use of the Jacobi identity \eq{omjac} for \mb{\omega^{AB}} and
the associativity \eq{balg} of the product \mb{T(y_{A},y_{B},y)}.

\proofbox

\end{appendix}




\end{document}